\documentclass{article}
\usepackage[utf8]{inputenc}
\usepackage{booktabs}
\usepackage{graphicx}
\usepackage{caption}
\usepackage{subcaption}
\usepackage{geometry}
\usepackage{xcolor}
\usepackage{soul}
\usepackage[skip=4pt]{caption}
\usepackage{authblk}
\usepackage{fancyhdr}
\usepackage{filecontents}
\usepackage{lineno}
\usepackage{indentfirst}
\usepackage[labelfont=bf]{caption}
\usepackage{placeins}

\newcommand\blfootnote[1]{%
  \begingroup
  \renewcommand\thefootnote{}\footnote{#1}%
  \addtocounter{footnote}{-1}%
  \endgroup
}

\newgeometry{vmargin={25mm}, hmargin={25mm,25mm}}   % set the margins
\begin{document}
\begin{figure*}[!h]  % spans both columns
\centering
\includegraphics[width=\linewidth,clip]{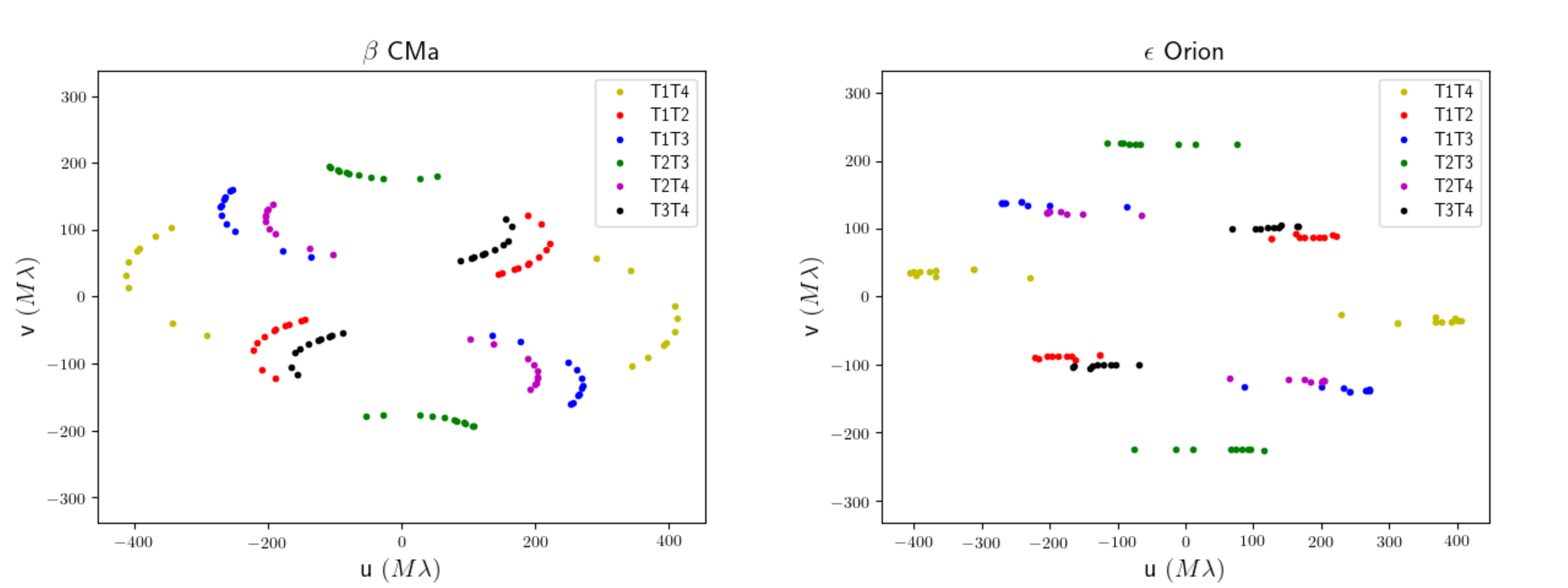}
\caption{\textbf{Coverage of the sources in the (u,v)-plane.} Each of the colored points represent different runs for a given telescope pair.}
\label{fig:uv_coverage}
\end{figure*}

%TC:ignore
\noindent \textbf{Demonstration of stellar intensity interferometry with the four VERITAS telescopes} \\ \\
\noindent
A.~U.~Abeysekara$^1$, W.~Benbow$^2$, A.~Brill$^{3}$,
J.~H.~Buckley$^{4}$, J.~L.~Christiansen$^{5}$, A.~J.~Chromey$^{6}$, M.~K.~Daniel$^{2}$, J.~Davis$^{1}$, A.~Falcone$^{7}$, Q.~Feng$^{3}$, J.~P.~Finley$^{8}$, L.~Fortson$^{9}$, A.~Furniss$^{10}$, A.~Gent$^{11}$, C.~Giuri$^{12}$, O.~Gueta$^{12}$, D.~Hanna$^{13}$, T.~Hassan$^{12}$, O.~Hervet$^{14}$, J.~Holder$^{15}$, G.~Hughes$^{2}$, T.~B.~Humensky$^{3}$, P.~Kaaret$^{16}$, M.~Kertzman$^{17}$, D.~Kieda$^{1,*}$, F.~Krennrich$^{6}$, S.~Kumar$^{13}$, T.~LeBohec$^{1}$, T.~T.Y.~Lin$^{13}$, M.~Lundy$^{13}$, G.~Maier$^{12}$, N.~Matthews$^{1,*}$, P.~Moriarty$^{18}$, R.~Mukherjee$^{19}$ M.~Nievas-Rosillo$^{12}$, S.~O'Brien$^{13}$, R.~A.~Ong$^{20}$, A.~N.~Otte$^{11}$, K.~Pfrang$^{12}$, M.~Pohl$^{21,12}$, R.~R.~Prado$^{12}$, E.~Pueschel$^{12}$, J.~Quinn$^{22}$, K.~Ragan$^{13}$, P.~T.~Reynolds$^{23}$, D.~Ribeiro$^{3}$, G.~T.~Richards$^{15}$, E.~Roache$^{2}$, J.~L.~Ryan$^{20}$, M.~Santander$^{24}$, G.~H.~Sembroski$^{8}$,
S.~P.~Wakely$^{25}$,
A.~Weinstein$^{6}$,
P.~Wilcox$^{9}$, D.~A.~Williams$^{14}$, T.~J~Williamson$^{15}$ \\
\blfootnote{$^{*}$ email: nolankmatthews@gmail.com; dave.kieda@physics.utah.edu
}
%\vspace{-10mm}
%\input{author_list}
%\textbf{Authors:} N. Matthews et. al., VERITAS Collaboration, T. LeBohec, J. Davis (placeholder) \\
\\
{\small
$^{1}$Department of Physics and Astronomy, University of Utah, Salt Lake City, UT 84112, USA \\ \\
$^{2}$Center for Astrophysics $|$ Harvard \& Smithsonian, Cambridge, MA 02138, USA \\ \\
$^{3}$Physics Department, Columbia University, New York, NY 10027, USA \\ \\
$^{4}$Department of Physics, Washington University, St. Louis, MO 63130, USA \\ \\
$^{5}$Physics Department, California Polytechnic State University, San Luis Obispo, CA 94307, USA \\ \\
$^{6}$Department of Physics and Astronomy, Iowa State University, Ames, IA 50011, USA \\ \\
$^{7}$Department of Astronomy and Astrophysics, 525 Davey Lab, Pennsylvania State University, University Park, PA 16802, USA \\ \\
$^{8}$Department of Physics and Astronomy, Purdue University, West Lafayette, IN 47907, USA \\ \\
$^{9}$School of Physics and Astronomy, University of Minnesota, Minneapolis, MN 55455, USA \\ \\
$^{10}$Department of Physics, California State University - East Bay, Hayward, CA 94542, USA \\ \\
$^{11}$School of Physics and Center for Relativistic Astrophysics, Georgia Institute of Technology, 837 State Street NW, Atlanta, GA 30332-0430 \\ \\
$^{12}$DESY, Platanenallee 6, 15738 Zeuthen, Germany \\ \\
$^{13}$Physics Department, McGill University, Montreal, QC H3A 2T8, Canada \\ \\
$^{14}$Santa Cruz Institute for Particle Physics and Department of Physics, University of California, Santa Cruz, CA 95064, USA \\ \\
$^{15}$Department of Physics and Astronomy and the Bartol Research Institute, University of Delaware, Newark, DE 19716, USA \\ \\
$^{16}$Department of Physics and Astronomy, University of Iowa, Van Allen Hall, Iowa City, IA 52242, USA \\ \\
$^{17}$Department of Physics and Astronomy, DePauw University, Greencastle, IN 46135-0037, USA \\ \\
$^{18}$School of Physics, National University of Ireland Galway, University Road, Galway, Ireland \\ \\
$^{19}$Department of Physics and Astronomy, Barnard College, Columbia University, NY 10027, USA \\ \\
$^{20}$Department of Physics and Astronomy, University of California, Los Angeles, CA 90095, USA \\ \\
$^{21}$Institute of Physics and Astronomy, University of Potsdam, 14476 Potsdam-Golm, Germany \\ \\
$^{22}$School of Physics, University College Dublin, Belfield, Dublin 4, Ireland \\ \\
$^{23}$Department of Physical Sciences, Cork Institute of Technology, Bishopstown, Cork, Ireland \\ \\
$^{24}$Department of Physics and Astronomy, University of Alabama, Tuscaloosa, AL 35487, USA \\ \\
$^{25}$Enrico Fermi Institute, University of Chicago, Chicago, IL 60637, USA \\ \\
}
\noindent \textbf{High angular resolution observations at optical wavelengths provide valuable insights in stellar astrophysics~\cite{monnier2003,Ridgway2019}, directly measuring fundamental stellar parameters~\cite{Boyajian2013,Casagrande2014}, and probing stellar atmospheres, circumstellar disks~\cite{Kraus2012}, elongation of rapidly rotating stars~\cite{vanBelle2012}, and pulsations of Cepheid variable stars~\cite{Kervella2017}. The angular size of most stars are of order one milli-arcsecond or less, and to spatially resolve stellar disks and features at this scale requires an optical interferometer using an array of telescopes with baselines on the order of hundreds of meters. We report on the successful implementation of a stellar intensity interferometry system developed for the four VERITAS imaging atmospheric-Cherenkov telescopes. The system was used to measure the angular diameter of the two sub-mas stars $\beta$ Canis Majoris and $\epsilon$ Orionis with a precision better than 5\%. The system utilizes an off-line approach where starlight intensity fluctuations recorded at each telescope are correlated post-observation. The technique can be readily scaled onto tens to hundreds of telescopes, providing a capability that has proven technically challenging to current generation optical amplitude interferometry observatories. This work demonstrates the feasibility of performing astrophysical measurements with imaging atmospheric-Cherenkov telescope arrays as intensity interferometers and the promise for integrating an intensity interferometry system within future observatories such as the Cherenkov Telescope Array.} \\
\\
%TC:endignore
\indent High-angular resolution optical astronomy is often performed through optical amplitude interferometry (OAI), which measures the source spatial coherence function by observing the fringe visibility of interference patterns produced when light collected at separated telescopes is superimposed~\cite{Labeyrie2006}. One other approach is stellar intensity interferometry (SII) that instead uses correlations of star-light intensity fluctuations recorded with independent detectors on separated telescopes to measure the spatial coherence~\cite{HBT1957c}. SII was developed in the late 1950s~\cite{hbtbook} and resulted in the Narrabri Stellar Intensity Interferometer (NSII) that was used for observations from 1963 to 1974, providing the first definitive catalog of 32 stellar angular diameters~\cite{HBT1974}. Observations with the NSII were limited to only very bright stars with visual magnitudes less than 2.5. The sensitivity of an SII telescope is linearly proportional to the telescope area and detector efficiency, and inversely proportional to the square root of the time-resolution as shown in the Methods section (see Equation~\ref{eqn:signaltonoise}). The capabilities of the NSII were restricted by these factors and concurrent technical advancements in OAI provided substantial gains in the achievable limiting magnitude using much smaller telescopes, forestalling further developments in astronomical SII efforts. \\
\\
\indent SII has re-emerged as a viable technique for high-angular resolution astronomy primarily due to the potential of outfitting current and future large diameter ($> 10\,$m) telescope arrays with SII capabilities~\cite{LeBohec2006, ASU2}. A suitable SII instrument requires recording the star-light intensity with nanosecond level time-resolution and then correlating the intensities between telescope pairs. The optical path length only needs control with a precision no better than a few centimeters, determined by the light travel distance over a duration equal to the detector time resolution. The insensitivity to optical imperfections allows for SII instrumentation to be installed onto imaging atmospheric-Cherenkov telescope (IACT) arrays constructed for gamma-ray astronomy. With mirror diameters typically exceeding 10\,m, IACTs are among the largest astronomical light collectors. IACTs employ fast (f\,$\sim$\,1.0) optics with segmented mirrors to reduce cost, and so the large light collection area is achieved at the expense of imaging resolution, but they remain capable for performing rapid optical photometry~\cite{lacki2011} and SII measurements of stars that are several magnitudes fainter than those accessible to the NSII~\cite{janvida}. There are several practical arguments for a modern SII observatory that uniquely complement the current capabilities of OAI. The tolerance to path length fluctuations allows for observations at all optical wavelengths, even in the U and B photometric bands that are generally inaccessible to OAI observatories on account of the required mechanical precision at a fraction of a wavelength. Furthermore, the baselines between telescopes can be made arbitrarily large in order to probe unprecedented angular scales as small as tens of microarcseconds with kilometer length baselines. A significant advantage of SII is that the technique can be scaled up to an arbitrary number of telescopes since only digital electronic connections are required~\cite{dravins2015}, enabling an optical intensity interferometry observatory that operates in a comparable way to radio interferometry. These realizations have led to several recent experimental efforts towards a modern intensity interferometer \cite{weiss2018,zampieri2016,Tan2016,Zmija2020}, successful SII observations of coherent intensity fluctuations using two telescopes~\cite{guerin2018, Acciari2019}, and distance calibration to the luminous blue variable P Cyg \cite{rivet2020}. We improve upon these observations by directly measuring angular stellar diameters through fits to the squared visibility-baseline dependence with an SII system extended to four telescopes to provide six simultaneous baselines. \\
\\
\begin{figure}
    \centering
    \includegraphics[width=\linewidth]{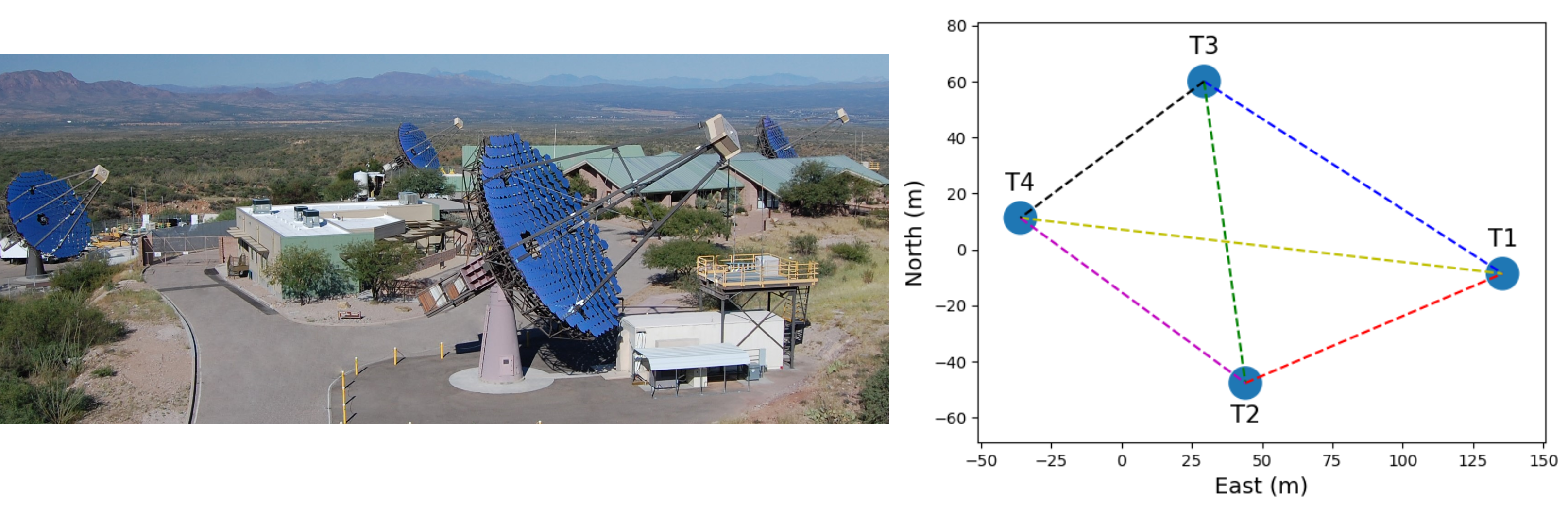}
    \caption{\textbf{The VERITAS array.} The left image shows a photograph of VERITAS located at the Fred Lawrence Whipple Observatory located in Amado, Az. The array consists of four 12\,m diameter telescopes, T1 (front-center), T2 (leftmost), T3 (rightmost), T4 (back-center). The right plot shows a top-down view of the array with each of the radial telescope separations.}
    \label{fig:veritas}
\end{figure}
\indent We report here on observations of the two bright ($m_v < 2.1$), hot (T $>22,000$\,K) B stars $\epsilon$\,Ori (Alnilam) and $\beta$\,CMa (Mirzam) that were conducted using an SII system (see Methods) installed onto the Very Energetic Radiation Imaging Telescope Array System (VERITAS) IACTs shown in Figure \ref{fig:veritas} ~\cite{kieda2019}. A total of 4.25 and 5.5 hours of data were taken of $\epsilon$\,Ori and $\beta$\,CMa, respectively, spanning the nights of December 12-14, 2019, local time. The interferometric (u,v)-plane coverage \cite{segransan2007} for both sources are shown in extended data Figure \ref{fig:uv_coverage}. The observations took place within a few days from the full moon when VERITAS does not operate as a gamma-ray instrument as the night sky background light overwhelms the faint Cherenkov signal. A custom camera is mounted near the focal plane of the VERITAS telescopes directly in front of the gamma-ray camera to enable SII observations. The starlight is passed through an interferometric filter with an effective center wavelength of $\lambda = 416\,$nm and bandpass of $\Delta \lambda = 13\,$nm that were chosen to match the peak reflectivity and quantum efficiency of the mirrors and detector. The filtered starlight is then detected by a photomultiplier tube. The resulting signal is continuously digitized and streamed to disk at a rate of 250\,MS/s at each telescope. Correlations between the intensities recorded at each telescope are performed off-line using a field-programmable gate array. We then analyze the correlated data by correcting for instrumental and geometrical time delays, background light effects, and applying noise cuts, to obtain the squared first-order coherence function $|g^{(1)} (\tau,\textbf{r}) |^2$ that is proportional to the squared visibility measured in OAI. It is a function of the path delay corrected time lag $\tau$ and projected baseline\,\,$\textbf{r}$ that is defined as the separation between the telescopes as viewed from the star. As the star tracks through the sky, a given telescope pair will span a range of projected baselines. The amplitude of $| g^{(1)} (\textbf{r}) |^2$ at $\tau =0$ measures the spatial coherence that is dependent on the source angular brightness distribution and projected baseline, reaching a minimum at the baseline $r \sim 1.22\lambda / \theta$ for an observation wavelength $\lambda$ and stellar angular diameter $\theta$. The angular diameter can then be determined by fitting the $|g^{(1)} (\textbf{r}) |^2$ measurements to an appropriate visibility model for a given source angular brightness distribution (see the Analysis section for more details). Stars that have a larger angular diameter will show a more rapid decline in $| g^{(1)} (\textbf{r}) |^2$ in comparison to stars with smaller angular diameters. \\

\begin{figure*}[t!]  % spans both columns
\centering
\includegraphics[width=\linewidth]{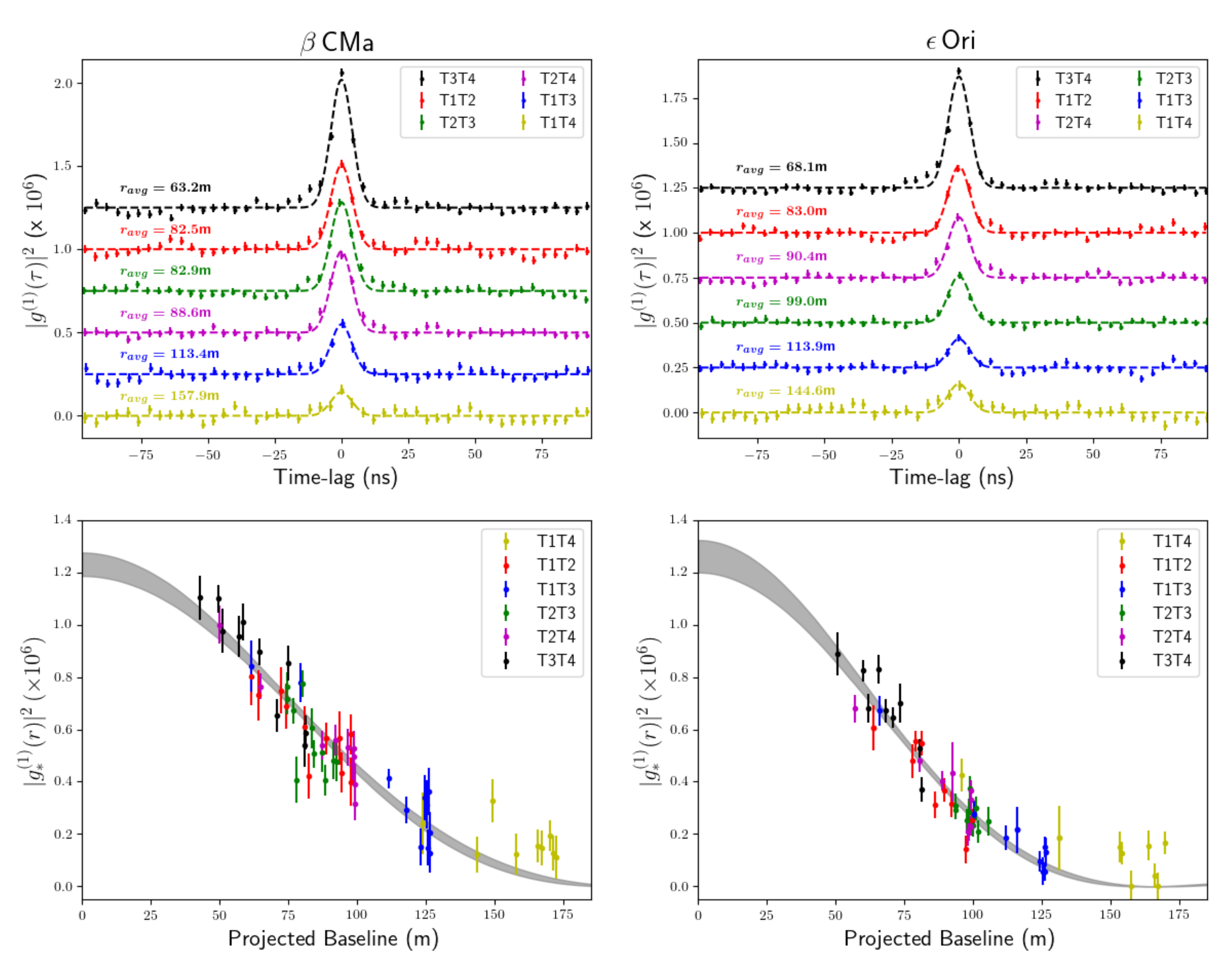}
\caption{\textbf{Temporal and spatial coherence measurements.} The top two panels show the averaged $|g^{(1)}(\tau)|^2$ correlation measurements over the full live time between different pairs of telescopes for $\beta\,$CMa (left) and $\epsilon$\,Ori (right). Each respective telescope pair measurement is color-coded matching the combinations shown in Figure \ref{fig:veritas}. The uncertainties are given by the standard deviation of the mean normalized correlation and are estimated independently for each time-lag. The correlations are ordered by increasing average baseline from the top, corresponding to decreasing spatial coherence. The dashed lines show Gaussian fits to the data. The amplitude and corresponding fit uncertainty of the $|g^{(1)}(\tau)|^2$ peak over shorter time intervals are obtained as a function of the baseline and shown by the individual points in the bottom panels. The uncertainty is determined by the square root of the covariance matrix resulting from the fit. These measurements are fit to a function that approximates the star as a uniform disk (see Eq. \ref{eqn:g2andvisb2}) and includes free parameters for the overall normalization and angular diameter. The shaded area shows the 68\,\% confidence intervals determined through the uncertainty in the fit parameters.}
\label{fig:mainresult}
\end{figure*}
Figure \ref{fig:mainresult} presents the temporal (top) and spatial (bottom) correlation measurements from the observations of $\beta\,$CMa (left) and $\epsilon$\,Ori (right). The top two panels show integrated $|g^{(1)} (\tau)|^2$ correlations for various pairs of the telescopes as a function of $\tau$. The significant peak at $\tau = 0$ is the signal associated with the spatial coherence of the star. The width of the $|g^{(1)} (\tau) |^2$ peak is determined by the correlation between the combined telescope optics and detector time response at each telescope. The dashed lines show Gaussian fits to the data. The Gaussian fit width, representative of the temporal resolution time of the system, is fixed to 4\,ns, as explained in Methods, and the amplitude is left as a free parameter. The lower panels of Figure\,\ref{fig:mainresult} show values of $|g_\ast^{(1)}(\textbf{r})|^2$ representing the correlation resulting only from star-light. These values are obtained by scaling the $|g^{(1)} (\tau,\textbf{r}) |^2$ amplitude fit by a factor that accounts for the effect of night sky background light and detector dark current. Each set of points are shown as a function of the projected baseline and correspond to an average of 17 minutes of data live-time per measurement. Using uniform-disk and limb-darkened models for the star angular brightness distribution, we perform a fit to the data with the stellar diameter and the zero-baseline correlation as free parameters. The shaded region shows the 68\,\% confidence regions obtained from the uniform-disk fit. The value of the zero-baseline correlation should be on the order of the ratio of the electronic to the optical bandwidth, each set by the instrumental system. In practice, the measured correlation can be affected by other factors including the telescope mirror extent, and spectral absorption or emission lines. We find that these additional effects are negligible in comparison to our measurement uncertainty, and thus the zero-baseline value should be the same for both stars. The zero-baseline normalization fit values are measured to be $N_0 = (1.23 \pm 0.05) \times 10^{-6}$ for $\beta$\,CMa, and $N_0 = (1.26 \pm 0.06) \times 10^{-6}$ for $\epsilon$\,Ori, consistent within fitted errors thus demonstrating the reliability of the system to systematic effects in observations of different stars. \\
\\
\indent For $\beta$\,CMa, we find a uniform disk diameter of $\theta_{UD} = 0.523 \pm 0.017\,$mas in agreement with the original NSII measurements of $\theta_{UD} = 0.50 \pm 0.03$\,mas \cite{HBT1974}. In the case of $\epsilon$ Ori, we obtain $\theta_{UD} = 0.631 \pm 0.017$\,mas, also in agreement with the NSII measurements ($\theta_{UD} = 0.67 \pm 0.04$\,mas). Using a limb-darkened model, given by Equation \ref{eqn:limbdarkenedmodel}, we find limb-darkened diameters of $\theta_{LD} = 0.542 \pm 0.018 $ for $\beta$\,CMa and $\theta_{LD} = 0.660 \pm 0.018$ for $\epsilon$\,Ori. We note that the VERITAS stellar intensity interferometer (VSII) gave more precise angular diameter measurements in comparison to the NSII, with less than a tenth of the observation time. The improvement is largely due to the greater light collection areas of the VERITAS mirrors, as well as the increased instantaneous baseline coverage offered by using multiple telescopes. To our knowledge, the angular diameters of these stars have not been measured since the time of the NSII. The current work thus provides confirmation of the NSII measurements of the stellar angular diameters to better than \,5\% uncertainty, a required capability for many science topics described later. Future observations can further reduce the uncertainty, as the root mean square fluctuations in the correlation should reduce inversely with the square root of the observation time. Here, we measure the uncertainty of the $|g^{(1)}(\tau)|^2$ fluctuations at a level of $ 2.0\times 10^{-8}$ for several telescope pairs using the entire data set for $\epsilon$\,Ori. These values correspond to an uncertainty in the squared visibility of 1.6\,\% for the measured zero-baseline correlation of $N_0 = 1.26 \times 10^{-6}$. Extrapolating these results to fainter stars demonstrates the current capability to measure squared visibilities with a precision of 4.0\,\% for stars of magnitude 2.5 and 10\,\% for stars with magnitude 3.5 for equivalent observation time. Direct comparison of these results with the NSII \cite{HBT1974} shows an improvement in the sensitivity by a factor of 6. Future improvements in the duty cycle and collimation of light through the narrowband filter offer expected gains in the sensitivity by a factor of 2 to 3, thus offering the potential for optical intensity interferometry measurements for stars with $m_B \sim 5$. \\
\\
\indent These two stellar angular diameter measurements with VSII demonstrate the feasibility of performing optical interferometry with IACT arrays. The approach developed here using high speed streaming with off-line correlations demonstrates the technical capability to perform optical intensity interferometry with tens to hundreds of telescopes, a capability that remains challenging for modern OAI observatories. We also show that SII observations increase the scientific output and scope of IACT observatories during bright moonlight conditions when gamma-ray observations are otherwise limited. The future Cherenkov Telescope Array (CTA) will employ up to 99 telescopes with up to kilometer baselines in the Southern hemisphere, and 19 telescopes in the Northern hemisphere with up to several hundred-meter baselines~\cite{acharyya2019}, allowing for unprecedented angular resolution capabilities approaching the tens of micro-arcsecond scale. The large number of telescopes would provide hundreds and possibly thousands of simultaneous baselines. Capability studies of SII on a CTA-like observatory demonstrate the ability to observe stellar targets brighter than a limiting visual magnitude of $m_V <$ 6 or 7~\cite{paul2012}. Science opportunities with a future CTA-SII observatory include surveying the angular diameter of stars larger than $\sim$ 0.05\, mas at short visible wavelengths, measuring the orbital and stellar parameters of interacting binaries~\cite{dravins2013}, characterizing the effects of gravity darkening and rotational deformation of rapidly rotating stars~\cite{nunez2015}, and imaging of dark or hot star spots~\cite{paul2012}. The VSII observations presented here demonstrate the core requirements for such an observatory, thus providing a technological pathway in addition to performing SII observations with unprecedented sensitivity. Our results demonstrate the capability to attain squared visibilities at the level of a few percent and therefore can directly complement ongoing science campaigns pursued by current OAI observatories~\cite{Stee2017}.
\section*{Correspondence}

\noindent Correspondence and requests for materials should be addressed to N. Matthews (nolankmatthews@gmail.com) and D. Kieda (dave.kieda@physics.utah.edu). 

\section*{Acknowledgements}

\noindent This research is supported by grants from the U.S. Department of Energy Office of Science, the U.S. National Science Foundation and the Smithsonian Institution, by NSERC in Canada, and by the Helmholtz Association in Germany. We acknowledge the excellent work of the technical support staff at the Fred Lawrence Whipple Observatory and at the collaborating institutions for their maintenance of VERITAS and assistance with integrating the SII system. The authors gratefully acknowledge support from NSF Grants No. AST 1806262, PHY 0960242, and the University of Utah for the fabrication and commissioning of the VERITAS-SII instrumentation.\\
\\
This work is dedicated in memory of Paul Nu\~{n}ez.

\section*{Author Contributions}

\noindent All authors contributed to the operation of the VERITAS telescopes and the internal review of the manuscript. N.M. and D.K. developed the SII hardware used in these observations with assistance from M.D. and G.H. in the integration with VERITAS; N.M. and D.K. wrote the data acquisition software, N.M. developed the FPGA correlator software; N.M. proposed the observations; N.M. and T.H. reduced and analyzed the data; N.M., D.K., T. LeBohec, T.Lin., D.R., Q.F., J.D., R.Y.S., U.A., A.F. took the observations; N.M. wrote the paper with guidance from T. LeBohec., M.D., D.K., T.H., M.P., and J.R.; T. LeBohec and D.K. provided key insights throughout the entire development of this work.

\section*{Competing Interests}

\noindent The authors claim no competing financial interests

%\printbibliography

\section*{Methods}

\subsection*{Instrumentation}

\indent The VERITAS observatory at the Fred Lawrence Whipple Observatory, located in Amado, AZ,  consists of an array of four 12 meter diameter telescopes of the Davies-Cotton design. The primary mirrors of each telescope consist of 345 hexagonal mirror facets arranged in f/1 optics. SII hardware is mounted onto an aluminum plate that is installed in front of the VERITAS gamma-ray camera. Initial tests of the hardware were performed in the laboratory \cite{matthews2018a} and with on-sky tests at the \textit{StarBase}-Utah observatory \cite{matthews2018b}. The equipment was then scaled for use on the VERITAS telescopes with successful tests of correlated intensity fluctuations with two telescopes \cite{Matthews2019}. On the plate, a 45$^{\circ}$ mirror redirects the light from the primary onto an optical diaphragm with a diameter corresponding to approximately 0.1$^{\circ}$ on the sky. The light then passes through an interferometric filter (SEMROCK 420/5) with a vendor-specified center wavelength of 420$\,$nm and bandpass of 5\,nm. Since the light is not collimated onto the filter the effective bandpass is broadened due to the large angle of incident light of up to 26.6$\,^{\circ}$. Calculations established the resulting bandpass to be centered on a wavelength of 416 nm with an effective width of 13\,nm. This effect reduces the overall spectral density throughput, and correspondingly, the signal-to-noise by approximately a factor of 2. The light is then detected by a Hamamatsu R10650 photomultiplier tube (PMT) with a quantum efficiency of $\sim\,$30\,\% at the observing wavelength \cite{otte2011}. The gain of the PMT is controlled via battery-powered high voltage. The exact high voltage delivered to the PMT is set by the duty cycle of a pulse-width modulator that is connected to the high voltage supply via an optical fiber \cite{cardon2019}. The output current of the PMT is fed into a low-noise trans-impedance FEMTO HCA-200M-20K-C 200$\,$MHz pre-amplifier. The voltage output of the pre-amplifier drives a long triaxial cable that is connected to the data acquisition system. A National Instruments (NI) NI-5761 DC-coupled analog-to-digital (ADC) converter continuously digitizes the amplified PMT signal at 250 MS/s with 14-bit resolution per sample over a peak to peak voltage range of 1.23$\,$V. Digitized values are passed to a NI PXIe-7965R module which hosts a Virtex 5 SX95T field-programmable gate array (FPGA). The FPGA is programmed to downcast the sampled value to an 8-bit integer, and push the data through a first-in-first-out buffer and then streamed to a 12TB NI-8265 RAID disk array at a data rate of 250$\,$MB/s per telescope. The ADC sample clock for each telescope is referenced and phase-locked to a common external 10 MHz clock. To generate the reference clock, a centrally located White Rabbit (WR) Switch is used to distribute a 10MHz clock through optical fiber connections to a WR-Len module located near each of the data acquisition systems. The WR-Len module takes in the optical fiber signal, and generates the electronic 10$\,$MHz signal to which the ADC sample clock is referenced. \\
\\
After observations are completed, the correlation between the stored data of each telescope is processed off-line by a FPGA-based correlator. The FPGA is programmed to retrieve the mean intensities of each time stream $\langle I \rangle = \frac{1}{N} \sum_i^N{I(t_i)}$, for the sample $i$, and the cross correlation between each pair of data streams, c(k) = $\langle I_1(t_i) I_2(t_{i+k}) \rangle = \frac{1}{N} \sum_i^N I_1(t_i) I_2(t_{i+k}) $, where $k$ is a digital time-lag inserted between the two channels. The correlation is calculated over 64 time-lag channels, corresponding to time-lags of -128 to +124$\,$ns in steps of 4$\,$ns.

\subsection*{Analysis}
\indent From the outputs of the correlator, the second-order coherence function can be calculated by normalizing the cross correlation by the product of the mean intensities
\begin{equation}
g^{(2)}(\tau, \mathbf{r}) = \frac{\langle I_1(t) I_2 (\mathbf{r},t+\tau) \rangle}{\langle I_1(t) \rangle \langle I_2(t) \rangle}
\label{eqn:g2}
\end{equation}
where $\mathbf{r} = \mathbf{r_1} - \mathbf{r_2}$ is the projected baseline between the two telescopes, $\tau$ is the relative time lag, and the brackets indicate a statistical average in time assuming a stationary light source. For chaotic thermal light, such as that from a star, the second-order coherence function can be written \cite{foellmi2009} in terms of the first-order coherence $g^{(1)}(\tau,\textbf{r})$,
\begin{equation}
    g^{(2)} (\tau, \mathbf{r}) = 1 + |g^{(1)}(\tau,\mathbf{r})|^2.
\end{equation}
\indent The van Cittert-Zernike theorem states that at $\tau=0$, the measured first-order coherence is equivalent to the Fourier transform of the source angular intensity distribution in the sky \cite{born1980}.  For randomly polarized light, the squared modulus of the first-order coherence is reduced by a factor of 1/2. In the typical case where the detector resolution time $T$ is much longer than the light coherence time $\tau_c \sim 1 / \Delta \nu$, where $\Delta \nu$ is the optical bandwidth, it is reduced by another factor of $\sim \tau_c / T$. We then write the expected zero-baseline coherence as $N_0 = \epsilon \tau_c / 2T$, where $\epsilon$ is a correction factor that accounts for the shape of the detected light spectral density that sets the optical bandwidth and corresponding coherence time. The correction factor also includes potential losses in the coherence signal that, as detailed by Hanbury-Brown and Twiss~\cite{hbtbook}, are attributed to non-ideal properties of the filters, detectors and electronic readout system that may influence the effective electronic or optical bandwidth. Under a uniform disk approximation for a star of angular diameter $\theta_{UD}$, the squared first-order coherence can be written as
\begin{equation}
    |g^{(1)}_\ast (\tau=0, r)|^2 =  N_0 \left| 2 \frac{J_1 (x_{UD})}{x_{UD}} \right| ^2
\label{eqn:g2andvisb2}
\end{equation}
where $x_{UD} = \pi \theta_{UD} r/\lambda$, $|g_\ast^{(1)}|^2$ is the coherence due to starlight alone, i.e. $g_\ast^{(2)} = \langle I_{1_\ast}I_{2_\ast}\rangle / (\langle I_{1_\ast}\rangle\langle I_{2_\ast}\rangle)$ where $I_{1_{\ast}}$ and $I_{2_{\ast}}$ are the starlight intensities at detectors 1 and 2, $J_1$ is the Bessel function of the first kind, and $\lambda$ is the mean observational wavelength. A linear limb-darkened model~\cite{HanburyBrown1974b} was also fit to the data, where the expected squared first-order coherence is given by 
\begin{equation}
    |g^{(1)}_\ast (\tau=0, \textbf{r})|^2 = N_0 \left( \frac{1-u_{\lambda}}{2} + \frac{u_{\lambda}}{3}\right)^{-2}  \left( (1-u_{\lambda}) \frac{J_1 (x_{LD})}{x_{LD}} + u_{\lambda} \sqrt{\pi/2} \frac{J_{3/2}(x_{LD})}{x_{LD}^{3/2}} \right)^2
    \label{eqn:limbdarkenedmodel}
\end{equation}
where $u_{\lambda}$ is the wavelength-dependent linear limb-darkening coefficient, and $x_{LD} = \pi \theta_{LD} r/\lambda$, where $\theta_{LD}$ is the limb-darkened angular diameter.\\
\\
\indent The measured spatial coherence signal can be affected by detector dark current and background light. The total signal intensity can be written as a sum of the starlight and background light sources $I(t) = I_{\ast} + I_{bkg}$ and using equation \ref{eqn:g2} it is straightforward to derive the relationship
\begin{equation}
    |g_\ast^{(1)}|^2 = |g^{(1)}|^2 \times (1+\beta_1) (1+\beta_2)
    \label{eqn:g2_nsb}
\end{equation}
under the assumption that the fluctuations in the starlight-background and background-background terms are uncorrelated (i.e. $\langle \Delta I_{1_{\ast}} \Delta I_{2_{bkg}} \rangle = \langle \Delta I_{1_{bkg}} \Delta I_{2_{\ast}} \rangle = \langle \Delta I_{1_{bkg}} \Delta I_{2_{bkg}} \rangle = 0$, where $\Delta I$ is the AC component of the given source intensity such that $I(t) = \langle I \rangle + \Delta I(t)$. The terms $\beta_1$ and $\beta_2$ are the ratios of the background to starlight intensity $I_{bkg}/I_{\ast}$ for a given telescope.\\
\\
\indent Hanbury Brown and Twiss \cite{HBT1957c} showed that the expected signal to noise ratio of the correlation due to the spatial coherence of the source, under a first-order approximation, is given as
\begin{equation}
    S/N = A\,\alpha\,n\,|g^{(1)}(r)|^2\,\sqrt{\frac{\Delta f T_0}{2}}
    \label{eqn:signaltonoise}
\end{equation}
where $A$ is the telescope mirror area, $\alpha$ is the quantum efficiency of the detectors, $n$ is the spectral flux density of the source given in units of ph s$^{-1}$ m$^{-2}$ Hz$^{-1}$, $\Delta f$ is the electronic bandwidth set by the time-response of the combined optical and electronic system, and $T_0$ is the observation duration. \\
\\
\indent For a given run, typically 20 to 30 minutes in total duration, the value of $\left| g^{(1)} (\tau) \right|^2$ is calculated over accumulation frames of 1 second. Any $\left| g^{(1)} (\tau) \right|^2$ frames contaminated by correlated high-frequency radio pickup are removed. The Fourier transform of each frame is calculated to identify corrupted frames. If the power at the specific noise frequency is greater than the average power over all other frequencies by a predetermined threshold cut, the frame is removed. For a typical run, 30 - 40\,\% of the data is removed as a result of the electronic noise. Additionally, any frames are removed where the mean PMT current falls below 5 $\mu A$, or approximately one-half of the total current due to star light. These current losses are attributed to the attenuation of star light by clouds, or also non-perfect tracking that results in the collected star-light not centered onto the detectors. Each frame is shifted in time by the average number of samples to zero optical path delay to correct for the geometrical optical path-delay. After path-delay corrections and noise/current cuts are applied, all $|g^{(1)} (\tau) |^2$ frames are averaged together through a weighted mean to produce a final $| g^{(1)} (\tau)|^2$ frame for that run. The weights for each frame are given by $1/\sigma^2_{|g^{(1)}|^2}$ where $\sigma_{|g^{(1)}|^2} = \sigma_{\langle I_1 I_2 \rangle} / \left( \langle I_1 \rangle \langle I_2 \rangle \right)$, where $\sigma_{\langle I_1 I_2 \rangle}$ is the standard deviation of the correlation over all time-lag bins for a given frame. Provided sufficient throughput and spatial coherence, the corrected and averaged results reveal an excess peak at zero time lag due to the spatial coherence of the source. In our case where the detector resolution time is much greater than the coherence time of the light, the shape of the peak is determined by the cross-correlation of the telescope and detector time response with an amplitude proportional to the amount of spatial coherence. The Davies-Cotton mirror design of the VERITAS telescopes introduces an approximate 4 ns spread in the arrival time of photons that otherwise would arrive synchronously \cite{HOLDER2006}. A PMT detects the photons, with a characteristic single photo-electron pulse width of 3.7 ns and additional timing jitter on the order of 1\,ns. Simulations of these effects find that the $\left| g^{(1)} (\tau) \right|^2$ peak can be modeled as a Gaussian following the form     
\begin{equation}
f(\tau) = A e^{-\frac{1}{2}(\tau-\tau_0)^2/ \sigma_{\tau}^2} + C
\end{equation}with $A$, $\tau_0$, and $C$ left as fit parameters. $A$ is the amplitude of the coherence peak and directly measures $|g^{(1)}(\textbf{r})|^2$, $\tau_0$ is a parameter that accounts for a variable start time between the separated data acquisition systems, and is constrained within $|\tau| < 10$\,ns of zero-lag, and C is another correction parameter that subtracts off any drift in the mean level of the correlator. The value of the fit width is fixed to a value of $\sigma_{\tau} = 4.0\,$ns determined empirically from the data and consistent with expectations from simulations. The value of the width was obtained by allowing it to be a free parameter in the fit of the integrated $\left| g^{(1)} (\tau) \right|^2$ correlation measurements for runs showing a peak amplitude with a p value less than 3 $\times 10^{-7}$. The value of the fit width was stable within fit uncertainties for several runs demonstrating that no significant fluctuations in the relative timing of the acquisition systems are present. \\
\\
\indent The value of the fit amplitude and corresponding uncertainty are found for all runs to measure $\left| g^{(1)} (\tau = 0) \right|^2$ as a function of the projected baseline. The projected baseline is calculated from the telescope positions and source sky position at the time of observation. The value of the fit amplitude is then multiplied by the scaling factor described in Equation \ref{eqn:g2_nsb} to compensate for night sky background and detector dark current. The mean level of the background is found by pointing the telescope off source and recording the average intensity. These off runs are taken before and after each run on source. The background light corrected values of $|g^{(1)}_\ast|^2$ are then fit to a function with the form of Equation \ref{eqn:g2andvisb2} in order to obtain the uniform-disk normalization and stellar diameter. Values of the linear limb-darkening coefficient $u_{\lambda}$ were estimated by interpolating existing tables that calculate expected values of $u_{\lambda}$ from atmospheric parameters~\cite{claret2011}. Using the atmospheric values listed in Table \ref{table:atm_params}, for observations in the B band, a microturbulent velocity of 2\,km/s, and solar metallicity  we find values of $u_{\lambda}$ of 0.38 and 0.46 for $\beta$\,CMa and $\epsilon$\,Ori, respectively. 
\begin{table}[h!]
\centering
\begin{tabular}{cccccc}
\toprule
Source & Spectral Type & $B$ & $T_{eff}$ (K) & $log(g)$ (dex) & $v\sin(i)$ (km/s) \\
\midrule
$\beta$ CMa & BII/III & 1.73 & 24000 $\pm$ 500$^{a}$ & 3.43 $\pm$ 0.10$^{a}$ & 20 $\pm$ 7$^{a}$ \\
$\epsilon$ Ori & B0Ia & 1.51 & 27000 $\pm$ 1000$^{b}$ & 2.90 $\pm$ 0.02$^{b}$ & 65 $\pm$ 15$^{c}$ \\
\end{tabular}
\caption{Atmospheric parameters used to estimate values of the linear limb-darkening coefficient. The spectral type and B band magnitudes were obtained from SIMBAD. $^{a}$ Taken from reference~\cite{Levenhagen2006}, $^{b}$ Taken from reference~\cite{Crowther2006}, $^{c}$ Taken from reference~\cite{Abt2002}.}
\label{table:atm_params}
\end{table}

\FloatBarrier

\subsection*{Extended Data}
\FloatBarrier
\subsection*{Data Availability}

\noindent The data that support the plots within this paper and other findings of this study are available from the corresponding author upon reasonable request (N.M). Due to the extremely large datasets obtained during the observations, in excess of 40 TB, only the post-correlation data can reasonably be made available.

\subsection*{Code Availability}
\noindent The software used to analyze the data in this study is available upon request to the corresponding author (N.M.).

%%TC:endignore
\end{document}